\documentclass[aps,prx,twocolumn,superscriptaddress,nofootinbib]{revtex4-2}

\usepackage{amsmath}
\usepackage{graphicx}
\usepackage{amssymb}
\usepackage{color}
\usepackage{mathrsfs}
\usepackage{verbatim}
\usepackage[colorlinks,linkcolor=blue,urlcolor=black,citecolor=blue,anchorcolor=blue]{hyperref}
\usepackage[T1]{fontenc}
\usepackage{orcidlink}
\usepackage{float}
\usepackage[british]{babel}

\newcommand{\rket}[1]{|{#1})}

\newcommand{\one}{\mathbf{1}}
\newcommand{\D}{\mathbf{D}}
\newcommand{\eps}{\boldsymbol{\epsilon}}
\newcommand{\rb}{\mathbf{r}}
\newcommand{\E}{\mathrm{E}}
\newcommand{\C}{\mathrm{C}}
\newcommand{\cm}{\mathrm{c}}
\newcommand{\Um}{\mathrm{U}}
\newcommand{\Sm}{\mathrm{S}}
\newcommand{\Rm}{\mathrm{R}}
\newcommand{\I}{\mathrm{I}}
\newcommand{\Bm}{\mathrm{B}}
\newcommand{\Lm}{\mathrm{L}}
\newcommand{\Mm}{\mathrm{M}}
\newcommand{\Z}{\mathbb{Z}}
\newcommand{\R}{\mathbb{R}}

\begin{document}

\title{Lattice models of chiral random motion: A short review}

\author{Erik Kalz\orcidlink{0000-0003-3294-7365}}
\affiliation{Universit{\'e} de Strasbourg, Institut de Sciences et d'Ing{\'e}nierie
Supramol{\'e}culaires, 67000 Strasbourg, France}
\email{erikkalz98@gmail.com}
\author{Jan W{\'o}jcik\orcidlink{0000-0001-7287-2244}}
\affiliation{University of Gda{\'n}sk, Institute of Theoretical Physics \& Astrophysics,
80-308 Gda{\'n}sk, Poland}
\author{Joshua Uhlig\orcidlink{0009-0007-5945-7020}}
\affiliation{University of Potsdam, Institute of Physics \& Astronomy, 14476 Potsdam, Germany}
\author{Nir Sherf\orcidlink{0009-0003-1031-4062}}
\affiliation{Tel Aviv University, School of Chemistry, Tel Aviv 6997801, Israel}
\author{Yael Roichman\orcidlink{0000-0003-1927-4506}}
\affiliation{Tel Aviv University, School of Chemistry, Tel Aviv 6997801, Israel}
\affiliation{Tel Aviv University, School of Physics and Astronomy, Tel Aviv 6997801, Israel}
\affiliation{Tel Aviv University, The Center for Physics and Chemistry of Living Systems,
Tel Aviv 6997801, Israel}
\author{Ralf Metzler\orcidlink{0000-0002-6013-7020}}
\email{ralf.metzler@uni-potsdam.de}
\affiliation{University of Potsdam, Institute of Physics \& Astronomy, 14476 Potsdam, Germany}

\date{\today}

\begin{abstract}
Chirality, a systematic rotational bias, is ubiquitous in many random motion patterns, from
bacteria cells, that swim in circles near surfaces and spinning colloids, to charged particles
diffusing under the influence of magnetic fields. In two dimensions, chirality renders random
transport \textit{odd}: the diffusion tensor acquires an antisymmetric part that leaves the
mean-squared displacement invariant but modifies fluxes and boundary currents. Lattice models
for random motion are a natural setting to study chirality, since they are exactly solvable,
connect directly to heterogeneous media, cellular automata, and programmable robots, and in
discrete time admit fully chiral limits with a localised bulk and edge transport. Lattice
walks with a rotational bias have been introduced in different communities several times since
1960, yet without a common language or conceptual connection between them. Here we review these
models from the perspective that on the lattice, chirality requires memory. Taking the recently
introduced chiral random walk model as a template, we describe the walker with an internal degree
of freedom. With a chiral biased updating rule, the full diffusion tensor of this chiral walker
turns odd. Within this picture, we can unite persistent and chiral walks, run-and-tumble motion,
and cyclic random motions in one framework. We discuss how the memory is followed into the
environment; in Lorentz lattice gases, kinetic and rotor walks, and random odd environments;
and into the path; in spiral self-avoiding walks.
\end{abstract}

\maketitle

\begin{center}
\textbf{Dedicated to Ewa Gudowska-Nowak}
\end{center}

\section{Introduction}
\label{sec:introduction}

In his 1888 work \emph{The Logic of Chance\/} \cite{venn}, John Venn writes: "I do not know that
any one has suggested the drawing of a line whose shape as well as position shall be of a
purely random character". Venn in fact presents a drawing of such random lines, in what we
would now call a random walk. The name "random walk" was coined later, by statistician Karl
Pearson in 1905 \cite{pearson}. Pearson's interest in such a process was triggered by his
colleague Ronald Ross, who was awarded the 1902 Nobel Prize for Physiology or Medicine
following his discovery on malaria transmission by mosquitoes. Ross described the basic
idea for understanding the spreading of infected mosquitoes from a breeding pool \cite{ross}:
"Suppose that a mosquito is born at a given point, and that during its life it wanders about,
to and fro, to left or to right, where it wills, in search of food or of mating, over a
country which is uniformly attractive and favourable to it. After a time it will die. What are
the probabilities that its dead body will be found at a given distance from its birthplace?"

Following Pearson's request for help, John William Strutt, Lord Rayleigh, provided the
solution that after many steps the location of the random walker is distributed according
to a Gaussian or normal law \cite{pearson,rayleigh}. In fact, the random walk is close to
Einstein's and Smoluchowski's formulations in 1905 and 1906 of the mathematical theory of
Brownian motion \cite{einstein,smoluchowski,ewabook}. Today, random walks are an almost
ubiquitous and extremely versatile tool to quantify the random spreading of entities ranging
from individual molecules to animals, to model polymer chains, or to capture the stochastic
dynamics of financial markets \cite{weiss,hughes,bouchaud,richter,benhamou,doi}. Direct
generalisations of the simple random walk are used in the modelling of anomalous diffusion
\cite{scher,brian,report,pccp,andi,igor,felix,yasmine,henrik,sandev,lenzi}, errors in telephone
communication \cite{mandelbrot,lewis}, persistent polymer models for brain fibre growth
\cite{skirmantas,skirmantas1,fibres-ou}, or to capture non-Gaussian distributions and
"roughness" in financial data \cite{bouchaud,stanley,comte,gatheral}.

In the "annus mirabilis" 1905, another breakthrough was reported, this time in medicine.
Fritz Schaudinn and Erich Hoffmann discovered that the bacteria \emph{Treponema pallidum\/}
is the causative agent of syphilis. In their work, however, they also describe how the
helically shaped cell body actively twists and spirals in fluids \cite{schaudinn,hofmann}.
Today, the interest in spiralling motion of bacteria and other single cells as well as
higher organisms is unbroken, ranging from the role of spirals in the high-speed motion
of the bacterium \emph{Magnetococcus marinus\/} \cite{bente}, the emerging swirling
dynamics of populations of chiral cells \cite{lichen}, to the motion of parasites such
as \emph{Trypanosoma brucei\/} \cite{stark}. One of the current research questions is
whether chirality might have a beneficial role in the motion of small active particles
such as bacteria or amoebae in narrow spaces such as soils or brain tissue, see, e.g.,
\cite{datta,datta1,alexandre} and references therein.

Concrete experimental observations of chiral motion are reported from systems across widely
different scales. Bacteria swim in circles near solid boundaries \cite{lauga2006swimming},
and asymmetric self-propelled colloids follow circular trajectories on substrates
\cite{kummel2013circular}. At the collective level, spinning colloids form chiral fluids
with persistent edge flows and odd hydrodynamic response \cite{soni2019odd}. More directly,
odd transport has now been resolved experimentally at the particle level. In a chiral active
bath, collisions transfer chirality to an embedded passive tracer, producing circular motion
and, under an applied force, a systematic transverse drift \cite{goerlich2026chirality}.
Odd diffusion itself has been measured in a two-dimensional fluid of air-driven chiral rotors,
for which the antisymmetric diffusion coefficient changes sign with the vorticity
\cite{vegareyes2022diffusive}. These experiments illustrate why chirality can remain hidden
in conventional measures of spreading: as argued below, the antisymmetric part of the
diffusion tensor does not enter the mean-squared displacement. Instead, effects of chirality
appear in cross-correlations, fluxes, boundaries, collisions and the inherent transverse
response.

Here, we are interested in the combination of random motion and chirality, specifically
in a quantitative description of chiral random walks in a two-dimensional setting and
its connection to "odd" diffusion and mobility. We provide an overview on the history
of lattice walks with a rotational bias and propose a common language for this variety
of models in terms of their memory, that we argue is the essential common denominator.
Specifically, we present the \emph{chiral random walk\/} (CRW) model introduced by us
recently.

In two dimensions, chirality manifests itself in antisymmetric off-diagonal elements of
transport tensors, and thus, interestingly, does not interfere with the isotropy of the
walk, but rather represents the most general form of isotropic transport tensors in two
dimensions \cite{dresselhaus2025anomalous}. This class of systems is also known as
"odd" \cite{fruchart2023odd} for the reason which will become clear soon. The most
general isotropic diffusion tensor in two dimensions, the "odd diffusion tensor"
reads \cite{hargus2021odd,kalz2022collisions}
\begin{equation}
\D=D_\parallel\one+D_\perp\eps=D_\parallel(\one+\kappa\eps),
\end{equation}
where $\one$ is the identity matrix, $\eps$ is the totally antisymmetric Levi-Civita symbol
($\epsilon_{xy}=-\epsilon_{yx}=1$ and zero otherwise), and $\kappa=D_\perp/D_\parallel$ the
"oddness parameter". The "odd diffusivity" $D_\perp$ typically is directly related to
the microscopic chirality and changes sign under parity or time reversal \cite{fruchart2023odd},
thus behaving odd under the inversion of chirality, which is the namesake for such 
odd systems. Note that observables that are based on positional arguments only are
insensitive to the odd diffusivity \cite{park2026role}. This can be intuitively understood
from the observation that the odd diffusivity drops out of the diffusion term $\nabla\cdot
(\D\nabla P)=D_\parallel\nabla^2 P$ in the diffusion equation, such that it leaves no trace
in the mean-squared displacement (MSD) of a free particle. It rather manifests itself through
flux-dependent observables, at boundaries and in collisions with obstacles or other particles 
\cite{kalz2022collisions}.

This invisibility explains a curious historical fact. Lattice walks with a preferred sense of
rotation have been introduced many times, yet the umbrella terminology to organise this
literature and to identify the common theme was missing---the concept of "odd transport"
now provides such an umbrella. Since transport typically was characterised through the MSD, the
odd part of such lattice walks simply remained unnoticed. We here organise these models around
the central idea: \textit{on the lattice, chirality requires memory}, and show that the
historically widely scattered lattice walks with chirality differ mainly in \textit{what
this memory is associated with\/}: to the walker itself or to the environment. The recently
introduced CRW model \cite{wojcik2026chiral} serves as our template, which essentially equips
the walker with an internal degree of freedom and thus stores two-step dynamical information.

This review is structured as follows: Section \ref{sec:memory} develops the language of
oddness and chirality and presents our central hypothesis that emergence of chirality requires
memory. Section \ref{sec:crw} specifically introduces the CRW model as a special case. Section
\ref{sec:walker} groups those works where the chirality sits "within" the walker, and section 
\ref{sec:environment} those where chirality is a property of the environment. We close in
section \ref{sec:outlook} with an outlook on contemporary literature.

\section{Why chirality requires memory}
\label{sec:memory}

We consider a random walker on the square lattice $(\ell\Z)^2$, with lattice constant $\ell$,
that carries an \emph{internal orientation\/} $d\in\mathcal{D}=\{\rightarrow,\downarrow,
\leftarrow,\uparrow\}$, such that the space $\mathcal{D}$ is  ordered clockwise. The associated
$z=4$ unit-displacement vectors $\mathbf{e}_d=\mathrm{E}|d)$ are collected in the \textit{lattice
geometry matrix\/} $\mathrm{E}$\footnote{We follow the convention of Ref.~\cite{sherf2026active}
that differs from the original model of Ref.~\cite{wojcik2026chiral} only by a permutation of
$\mathcal{D}$.}, $z$ being the coordination number of the lattice. The joint probabilities of
position and internal degree, $P_d(\rb,t)$ constitute the vector $|\rho(t))=\sum_{\rb,d}P_d(\rb,
t)\,|\rb)\otimes|d)$, where the product state $|\mathbf{r})\otimes|d)\equiv|\mathbf{r},d)$ resides
in the product space $(\ell\Z)^2\otimes\mathcal{D}$. The full time evolution acts on the product
state and evolves the probability vector in time as
\begin{equation}
\label{eq:CRW_time_evolution}
|\rho(t+1))=\Sm\left(\sum_{\rb}|\rb)(\rb|\otimes\C_\mathbf{r}\right)|\rho(t)),
\end{equation}
Each step consists of two operations: The coin $\C_{\rb}$ is a $z\times z$ stochastic matrix
that acts on the orientation only---its element $(d'|\C_{\rb}|d)$ is the probability that a
walker at site $\rb$ with orientation $d$ adopts the new orientation $d'$, so that $\sum_{d'}
(d'|\C_{\rb}|d)=1$. The conditional shift $\Sm=\sum_{\rb,d}|\rb+\ell\mathbf{e}_d)(\rb|\otimes
|d)(d|$ then moves the walker by one lattice spacing along its new orientation. The index
$\rb$ allows the coin to vary in space, to account for heterogeneous environments
\cite{uhlig2026chirality}, see section \ref{sec:environment}. In an isotropic, homogeneous
system, the coin is identical everywhere, $\C_{\rb}=\C$, and the one-step operator reduces to
$\Um=\Sm(\one\otimes\C)$, which we use until section \ref{sec:environment}. This coin-shift
structure is borrowed from discrete-time quantum walks \cite{Kempe2003quantum}, with the
unitary coin replaced by a stochastic coin, and interestingly allows us to sort the historical
literature on chiral walks.

For a homogeneous, unbiased walk with stationary (orientation) distribution $\boldsymbol{\pi}$,
satisfying $\mathrm{C}\boldsymbol{\pi}=\boldsymbol{\pi}$ and writing $\Pi=\mathrm{diag}(
\boldsymbol{\pi})$, we can write the long-time diffusion tensor from the discrete Green-Kubo
relation \cite{wojcik2026chiral,gilbert2010diffusion} as
\begin{equation}
\label{eq:odd_green_kubo}
\D=\frac{\ell^2}{\Delta t}\E\left(\frac{1}{2}\one+\sum_{t=1}^\infty\C^t\right)\Pi\E^\mathrm{T},
\end{equation}
where the $t$th term of the sum is the correlation between a step and the $t$th subsequent
step, $\langle\mathbf{e}_{d_{s+t}}\otimes\mathbf{e}_{d_s}\rangle$, and $\mathbf{e}_{d_s}$ is
the direction at time $s$. The flux of the position density $\rho(\rb,t)=\sum_dP_d(\rb,t)$, the
marginal of $|\rho(t))$ over the orientation, thus reads $\mathbf{j}=-\D\nabla\rho$, and only
the symmetric part of $\D$ governs the spreading.

Equation \eqref{eq:odd_green_kubo} then allows us to state the central hypothesis of this
review more precisely: \textit{on the lattice, chirality requires memory}, i.e., an odd
diffusion tensor requires that the next step depends at least on the previous step, in a way
that is not invariant under time reversal at fixed parameters. Three observations follow from
this hypothesis:

\begin{itemize}

\item[(i)] If the steps of the walker are uncorrelated, $\mathrm C=\pi\one^\mathrm{T}$
\cite{feller1971introduction}, the increments are independent and all terms with $t\geq1$
vanish. $\D=(\ell^2/2\Delta t)\,\E\Pi E^\mathrm{T}$ then is symmetric: without memory, the
walk is not odd.

\item[(ii)] If the orientational dynamic obeys detailed balance at fixed parameters, $\C\Pi=
\Pi\C^\mathrm{T}$ \cite{norris1998markov}, then $\E\C^t\Pi\E^\mathrm{T}=(\E\C^t\Pi\E^\mathrm{
T})^\mathrm{T}$ for all $t$, and $\D$ is symmetric\footnote{The same holds when the detailed
balance condition is combined with the reversal of all steps, $\C\Pi=\Pi\Bm\C^{\mathrm{T}}\Bm$,
with $\Bm|d)=|{-d})$ reversing the orientation, since $\E\Bm=-\E$.}: memory alone is not
sufficient, it must also break time-reversal symmetry.

\item[(ii$^*$)] If instead the time-reversed dynamic coincides with the dynamic in which the
source of chirality is reversed\footnote{For the isotropic coins of equation
(\ref{eq:geneal_coin}), $\tilde\C$ follows from $\C$ by exchanging $q_L$ and $q_R$.}, and thus
falls in the class of the Onsager-Casimir symmetry \cite{casimir1945onsager}, then $\C\Pi=\Pi
\tilde\C^{\mathrm{T}}$ and an odd part of the diffusion tensor is allowed, $\D(\C)=\D(\tilde{
\C})^{\mathrm{T}}$. 

\end{itemize}

\begin{figure}
\includegraphics[width=\columnwidth]{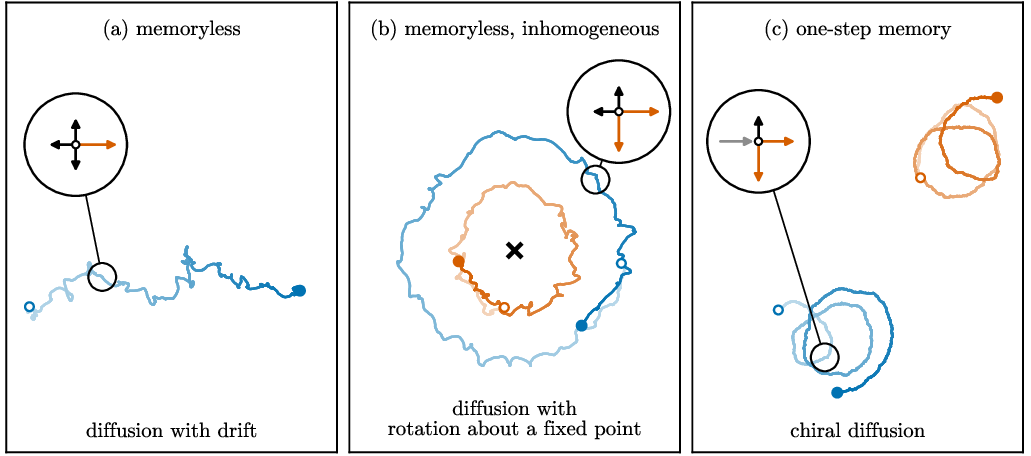}
\caption{Illustration of memory in a random walk. Open dots mark the initial positions, closed
dots the final positions of the shown trajectories. Magnified insets show the step probabilities,
with orange arrows for the more probable directions. (a) A memoryless walker whose bias is fixed
in space diffuses with a systematic drift, a polar response without rotation. (b) A memoryless
walker whose bias refers to a fixed point ($\times$), here tangential, circulates about that
point: without memory, rotation requires an inhomogeneous environment that supplies the centre,
cf.\ figure \ref{fig:environment}(c). (c) A walker whose bias refers to its previous step (grey
arrow), here favouring straight and right steps as for the persistent chiral coin of the active
CRW, circles about its own, moving centre and diffuses chirally.}
  \label{fig:memory}
\end{figure}

\begin{figure}
\includegraphics[width=\columnwidth]{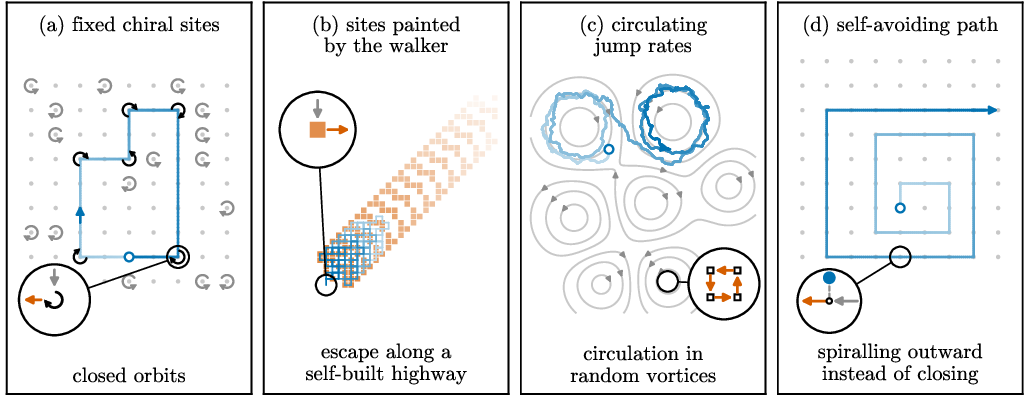}
\caption{Memory in the environment and in the path. (a) Fixed chiral sites: in a Lorentz
lattice gas of rotators with empty sites (curved arrows, black when visited), the particle
moves ballistically and turns at every rotator, here on a closed orbit; the inset shows a
right-turning rotator. (b) Sites painted by the walker: Langton's ant turns right on white
and left on painted (orange) squares and flips their colour; after a long transient it
escapes along a periodic highway that it builds itself. (c) Circulating jump rates: a
walker without internal state, whose jump rates carry an antisymmetric part given by the
lattice curl of a random stream function (grey contours), circulates in the resulting random
vortices; the inset shows the circulating rates around an elementary square. (d) Self-avoiding
path: a spiral self-avoiding walk steps only straight or to the right and never is allowed to
revisit a site; where a right turn would hit its own path (blue dot in the inset), it has to
go straight, so that it winds outward instead of closing.}
\label{fig:environment}
\end{figure}

For homogeneous walks, observations (i) and (ii) prove the hypothesis. When the memory resides
in the environment or in the path (section \ref{sec:environment}), however, it serves as our
organising principle. Observation (ii$^*$) reflects the Onsager-Casimir reciprocity of equilibrium 
systems whose parameters are odd under time reversal, such as a charged Brownian particle in a
magnetic field \cite{casimir1945onsager}, which obeys equilibrium dynamics
\cite{balakrishnan2021elements} and is odd \cite{chun2018emergence}. Whether a chiral walk
describes an equilibrium or an active system is therefore decided by whether its source of
chirality is reversed under time reversal, such as a magnetic field, or not, such as an
active torque. A walker governed by equation \eqref{eq:CRW_time_evolution} stores the required
memory in its orientation; it may equally reside in the environment or in the path of the
walker, see figures \ref{fig:memory} and \ref{fig:environment}.

We start with the elementary motions on the square lattice, with coordination number $z=4$.
Labelling the orientations in clockwise order as specified above, $d_0={\rightarrow}$,
$d_1={\downarrow}$, $d_2={\leftarrow}$ and $d_3={\uparrow}$, we define the rotation $\Rm|d_m)=
|d_{m+1})$, with indices modulo four, which turns the walker clockwise (right) by $\pi/2$; its
inverse $\Lm=\Rm^{-1}$ turns the walker counterclockwise (left). We call a coin isotropic if it
commutes with this rotation, $\Rm\C=\C\Rm$, i.e., if the probability of each turn does not depend
on the current orientation. Isotropy here refers to rotations only, since a reflection of the
lattice maps $\Lm$ onto $\Rm$. Because the four eigenvalues of $\Rm$ are distinct, every matrix
that commutes with $\Rm$ is a polynomial in $\Rm$, so that an isotropic coin is a combination of
the four elementary motions, persistence $\I=\Rm^0$, a right turn $\Rm$, reversal $\Bm=\Rm^2$,
and a left turn $\Lm=\Rm^3=\Rm^{-1}$ \cite{sherf2026active}.

By this definition the general coin matrix is an element of the group algebra of the cyclic
group $\C_4$, and it reads
\begin{equation}
\label{eq:geneal_coin}
\C=q_I\I+q_L\Lm+q_B\Bm+q_R\Rm+q_M\Mm,
\end{equation}
where the nonnegative probability weights $q_i$ sum to unity due to normalisation. We provide a
matrix representation of the elementary operators and the eigenvalues of the general coin in
appendix \ref{sec:matrices}. The mixing operator $\Mm=(\I+\Lm+\Bm+\Rm)/4=\boldsymbol{\pi}\one
^\mathrm{T}$ is randomising the orientations and represents the memoryless step of observation
(i) above. The physical interpretation of the weights $q_i$ can be elucidated by projecting the
$z=4$-dimensional coin operator onto the physical lattice space \cite{sherf2026active}, observing
that $\E\C=\cm\E$ where $\cm=q_p\one+q_c\eps$ is the two-dimensional (spatial) representation of 
the coin. Here $q_p=q_I-q_B$ measures the "persistence", i.e., the excess of forward over backward
steps and $q_c=q_R-q_L$ quantifies the chirality, i.e., the excess of right over left steps.

If we define the complex parameter $\lambda=q_p+\mathrm{i}q_c$, equation \eqref{eq:odd_green_kubo}
becomes a geometric series in $\cm$, yielding $\D=(\ell^2/4\Delta t)(\one+\cm)(\one-\cm)^{-1}$;
explicitly,
\begin{subequations}
\label{eq:odd_diff_general_coin}
\begin{align}
\D&=\frac{\ell^2}{4\Delta t}\left(\frac{1-|\lambda|^2}{|1-\lambda|^2}\one+\frac{2\,\mathrm{Im}
(\lambda)}{|1-\lambda|^2}\eps\right)\\
&=\frac{\ell^2}{4\Delta t}\left(\frac{1-q_p^2-q_c^2}{(1-q_p)^2+q_c^2}\one+\frac{2q_c}{(1-q_p)^2
+q_c^2}\eps\right).
\end{align}
\end{subequations}
The odd diffusion coefficient $D_\perp\propto q_c$ is thus induced by the chirality $q_c$ of
the walk. Since isotropic coins are doubly-stochastic, time reversal maps $\C$ onto $\C^
\mathrm{T}$, which exchanges $\Lm$ and $\Rm$ and thus reverses $q_c$; a reflection $\mathrm{P}$
of the lattice does the same, $\mathrm{P}\C\mathrm{P}=\C^\mathrm{T}$. A chiral coin, $q_c\neq0$,
therefore breaks parity and time-reversal symmetry separately (but not their combination) and
produces odd diffusion; if its source of chirality is reversed under time reversal, as in
observation (ii$^*$), this happens at equilibrium.

We note that $\lambda$ is the combination of the four polar eigenvalues $1$, $\mathrm{i}$, $-1$,
and $-\mathrm{i}$ of $\I$, $\Rm$, $\Bm$, and $\Lm$ weighted by the respective coin probabilities,
so that all coin operators map onto the square $|q_p|+|q_c|\leq1$ \cite{sherf2026active}. On
lattices with a coordination number of $z\geq 3$ equivalent directions, the elementary motions
are rotations of the cyclic group $C_z$, and $\lambda$ lies in the regular polygon spanned by
the $z$th roots of unity, and equation \eqref{eq:odd_diff_general_coin} holds without change.
For $z\to\infty$, $\lambda$ fills the unit disk. 

Equivalently, $q_p=\langle\cos\Delta\theta\rangle$ and $q_c=\langle\sin\Delta\theta\rangle$ are
the mean cosine and sine of the turning angle $\Delta\theta$, so that the complex parameter 
$\lambda=q_p+\mathrm{i}q_c=\langle\mathrm{exp}(\mathrm{i}\Delta\theta)\rangle$ is the mean turning
phase. Note that this average can also be used to define chirality for lattices with arbitrary
angles.

\section{The chiral random walk}
\label{sec:crw}

The CRW model of Ref.~\cite{wojcik2026chiral} is a special case of the general form in equation
\eqref{eq:geneal_coin}. Namely, it interpolates between the memoryless and fully chiral coin,
$\C_\mathrm{CRW}=(1-\chi)\Mm+\chi\Rm$ with $0\leq\chi\leq1$, so that $\lambda=\mathrm{i}\chi$
and, from equation \eqref{eq:odd_diff_general_coin},
\begin{equation}
\label{eq:difufsion_CRW}
D_\parallel^\mathrm{CRW}=\frac{\ell^2}{4\Delta t}\frac{1-\chi^2}{1+\chi^2},\qquad D_\perp
^\mathrm{CRW}=\frac{\ell^2}{4\Delta t}\frac{2\chi}{1+\chi^2}.
\end{equation}
$D_\parallel^\mathrm{CRW}$ decreases
monotonically with $\chi$, for $\chi\to1$ the walker encircles predominantly a single elementary
square. In the limit $\chi=1$, the Fourier-transformed propagator $\hat{\Um}(\mathrm{q})=\hat{
\mathrm{S}}(\mathrm{q})\Rm$ with the step $\hat{\mathrm{S}}(\mathrm{q})=\mathrm{diag}(\exp(-
\mathrm{i}\ell\mathbf{q}\cdot\mathbf{e}_d))$ becomes unitary and satisfies $\hat{\Um}(\mathrm{q}
)^4=\one$, such that its spectrum is flat, and no dissipative bulk transport takes place any more,
formally $D_\parallel^\mathrm{CRW}=0$ in equation \eqref{eq:difufsion_CRW}. In
Ref.~\cite{wojcik2026chiral} this limit is identified with a topologically non-trivial
discrete-time quantum walk in which clockwise and counterclockwise walks belong to different
phases. The celebrated bulk-boundary correspondence for topologically protected systems holds,
as well: reflective boundaries cause an edge state. This phenomenon is preserved quantitatively
in the dissipative regime $\chi<1$, states initialised at the boundary decay significantly slower
than in the bulk. The CRW thus models and explains the robust boundary fluxes seen for
odd-diffusive systems \cite{hargus2021odd,fruchart2023odd} from a new perspective. 

This observation carries a phenomenological prediction: moderate heterogeneity in the
environment should be beneficial for odd-diffusive motion. In Ref.~\cite{uhlig2026chirality},
the CRW was therefore placed on site-percolating clusters. Accessible sites appear with
occupation probability $p$, and blocked sites are modelled with the reflective coin $\Bm$.
The annealed coin $\C_\mathrm{ann}=p\C_\mathrm{CRW}(\chi)+(1-p)\Bm$ has $\lambda=-(1-p)+
\mathrm{i}p\chi$, and equation \eqref{eq:odd_diff_general_coin} yields
\begin{eqnarray}
\nonumber
D_\parallel^\mathrm{ann}=\frac{\ell^2}{4\Delta t}\frac{p[2-(1+\chi^2)p]}{(2-p)^2+\chi^2p^2},\\
D_\perp^\mathrm{ann}=\frac{\ell^2}{4\Delta t}\frac{2p\chi}{(2-p)^2+\chi^2p^2}.
\label{eq:annealed_CRW}
\end{eqnarray}
At fixed $p$, this $D_\parallel^\mathrm{ann}$ decreases again monotonically with $\chi$. In
$p$, however, $D_\parallel^\mathrm{ann}$ is non-monotonic: uncorrelated reversals truncate
the chiral orbits, so that obstacles \emph{enhance\/} the diffusivity of a strongly chiral
walker above its obstacle-free value. In quenched media, Ref.~\cite{uhlig2026chirality}
reports, in contrast, an optimal chirality $\chi^*(p)$ which maximises diffusion at every
obstacle density above percolation by up to almost a factor of three. This effect is reached
when a chiral walker follows the wall for about half the typical perimeter of an obstacle
cluster, an effect of quenched correlations and reminiscent of the collision-enhanced
self-diffusion in odd-diffusive systems
\cite{kalz2022collisions,kalz2024oscillatory,langer2024dance}. 

So far, $\lambda$ was purely imaginary and thus chirality is coupled to the loss of
persistence. The active CRW of Ref.~\cite{sherf2026active} instead uses the full coin
matrix \eqref{eq:geneal_coin} and thus also mixes persistent motion with chirality. The
directional correlations $\langle\mathbf{e}_{d_{s+t}}\otimes\mathbf{e}_{d_{s}}\rangle$ now decay
as $|\lambda|^t$ and rotate by $\mathrm{arg}(\lambda)$ per step---and both hydrodynamic
coarse-graining and the Green-Kubo route reproduce equation \eqref{eq:odd_diff_general_coin}.
In the \textit{weak-turning\/} limit $\lambda\approx1-(D_r-\mathrm{i}\omega)\Delta t$, and
the active CRW reproduces the transport characteristics of the chiral active Brownian particle
model with rotational diffusion $D_r=(1-q_p)/\Delta t$ and active torque $\omega=q_c/\Delta t$
\cite{liebchen2022chiral}. Strong chirality and anti-persistence, $q_p<0$, however, remain
intrinsically discrete \cite{sherf2026active}. The topological chiral random walker of 
Ref.~\cite{osat2026topological} that was introduced around the same time as the CRW model of
Ref.~\cite{wojcik2026chiral} adds rotational-noise steps without translation and exhibits
topologically protected edge currents that allow it to solve mazes.

\section{Memory in the walker}
\label{sec:walker}

We here discuss memory from the perspectives of discrete time and continuous time settings.

\subsection{Discrete time: correlated and persistent walks}

Walks with the concept of an internal degree of freedom that keeps track of the direction of
motion date back to 1953 in the work of Patlak \cite{patlak1953random}, yet without a chiral
bias. The first explicit chiral lattice walk we are aware of is that by Gillis in 1960
\cite{gillis1960random}. This work computed exact return probabilities of a walker on the
square lattice that turns right or left at every step with probabilities $(1\pm\alpha)/2$, $-1
\leq\alpha\leq 1$. The associated coin, in our language, $[(1+\alpha)\Rm+(1-\alpha)\Lm]/2$
shares $\lambda=\mathrm{i}\alpha$ with the CRW, and thus has the same diffusion tensor,
including the odd part, as equation \eqref{eq:difufsion_CRW}. Kareiva and Shigesada
\cite{kareiva1983analyzing} then considered a walk with random step length $\ell$ and
continuous turning angles $\theta$, i.e., for $z\to\infty$. Here $\lambda=\langle\cos\theta
\rangle+\mathrm{i}\langle\sin\theta\rangle$ and the odd diffusion tensor follows in the form
\begin{eqnarray}
\nonumber
\D=\frac{1}{4\Delta t}\left[\langle\ell^2\rangle\one+2\langle\ell\rangle^2\,\cm(\one-\cm)^{-1}
\right],\\
\cm=\langle\cos\theta\rangle\,\one+\langle\sin\theta\rangle\,\eps,
\end{eqnarray}
which reduces to equation \eqref{eq:odd_diff_general_coin} for constant $\ell$. Lattice walks
with direction-dependent \cite{henderson1984correlated} and arbitrary \cite{chen1994general}
transition matrices fall into the realm of equation \eqref{eq:odd_green_kubo}, as well as
Larralde's explicitly chiral persistent walk \cite{larralde1997transport}, exact propagators
for asymmetric angular kernels \cite{bracher2004eigenfunction} and spiralling paths from 
asymmetric turning distributions \cite{sadjadi2015persistent} complete this lineage. Walks
that remember several previous steps \cite{gilbert2011diffusive,gilbert2010diffusion} carry
larger internal states, i.e., incorporate memory beyond two steps.

\subsection{Continuous time: chiral run-and-tumble and cyclic random motions}

In continuous time, the coin becomes a generator. If the walker applies $\Rm^k$, i.e., turns
by an angle $\theta_k=2\pi k/z$ with $1\leq k<z$, with probability $\gamma_k\Delta t$ per
step, and moves the distance $\ell=v\Delta t$, the coin is $\C_\mathrm{cont}=\I+\mathrm{Q}\,
\Delta t$. Here $\mathrm{Q}=\sum_k\gamma_k(\Rm^k-\I)$ is the generator for the orientational 
dynamics \cite{norris1998markov}. For $\Delta t\to0$, equation \eqref{eq:CRW_time_evolution}
becomes $\partial P_d/\partial t=-v\,\mathbf{e}_d\cdot\nabla P_d+\sum_{d'}Q_{dd'}P_{d'}$, 
a velocity-jump process with continuous positions and discrete headings \cite{othmer1988models}.
The polar eigenvalue of the coin is $\lambda=1-(\Gamma-\mathrm{i}\,\Omega)\Delta t$, with
decorrelation rate $\Gamma=\sum_k\gamma_k(1-\cos\theta_k)$ and mean angular velocity $\Omega=
\sum_k\gamma_k\sin\theta_k$. Equation \eqref{eq:odd_green_kubo} then turns into $\D=v^2\int_0
^\infty\mathrm{d}t\\E\,\mathrm{exp}(\mathrm{Q}t)\,\Pi\E^{\mathrm{T}}=(v^2/2)(\Gamma\one-\Omega
\eps)^{-1}$, i.e., we have
\begin{equation}
D_\parallel^\mathrm{cont}=\frac{v^2}{2}\,\frac{\Gamma}{\Gamma^2+\Omega^2},\qquad D_\perp
^\mathrm{cont}=\frac{v^2}{2}\,\frac{\Omega}{\Gamma^2+\Omega^2}.
\label{eq:ct}
\end{equation}
For four directions, $\Gamma=\gamma_L+2\gamma_B+\gamma_R$ and $\Omega=\gamma_R-\gamma_L$, with
$\gamma_{R,B,L}=\gamma_{1,2,3}$, reproduce the chiral random motion of Hargus \textit{et al.}
\cite{hargus2021odd} and the chiral run-and-tumble walker of Mallikarjun and Pal
\cite{mallikarjun2023chiral}, with orthogonal \cite{angelani2022orthogonal} and classical
\cite{martens2012probability} run-and-tumble walks as relatives. For $z\to\infty$\footnote{For
instance, with turns by the smallest angles $\pm\Delta\theta=\pm2\pi/z$ only, at rates $\gamma
_{R,L}=D_r/\Delta\theta^2\pm\omega/(2\Delta\theta)$, the heading performs drifted rotational
Brownian motion, with $\Gamma=2D_r(1-\cos\Delta\theta)/\Delta\theta^2\to D_r$ and $\Omega=\omega
\sin\Delta\theta/\Delta\theta\to\omega$.}, $\Gamma\to D_r$ and $\Omega\to\omega$ recover the
model of chiral active Brownian particles \cite{hargus2021odd,liebchen2016pattern}.

The oddness of these walks, $\kappa=D_\perp^\mathrm{cont}/D_\parallel^\mathrm{cont}=\Omega/
\Gamma$, compares how fast the heading rotates with how fast it decorrelates. A turn by a
rotation of the walker by an angle $\theta_k$ contributes $\gamma_k\sin\theta_k$ to the former
and $\gamma_k(1-\cos\theta_k)$ to the latter, and the magnitude of their ratio, $|\cot(\theta_k
/2)|$, is largest for the smallest turns the lattice allows, $\theta_k=\pm2\pi/z$. Hence
$|\kappa|\leq\cot(\pi/z)$, where the equality holds in the case of cyclic motion. Thus, in
continuous time, only the limit $z\to\infty$ allows for an unbounded $\kappa$, in contrast to
the case of discrete time, where the $\chi\to1$ limit of the CRW corresponds, formally, to
$\kappa\to\infty$ \cite{wojcik2026chiral}. Therefore, only discrete time on lattices
permits access to complete localisation and the edge states of section \ref{sec:crw}.

The fully chiral generator $\mathrm{Q}=\gamma(\Rm^{\pm1}-\I)$ defines the cyclic motions of
probability theory \cite{di2002exact,orsingher2002bessel,leorato2003alternating, 
leorato2004bose,orsingher2004cyclic,lachal2006cyclic,lachal2006minimal}. Since $\Rm^{\pm1}$
shifts the probability from $d_{m\mp1}$ to $d_m$, the velocity-jump equation becomes $A_{d_m}
P_{d_m}(\rb,t)=\gamma P_{d_{m\mp1}}(\rb,t)$ with $A_d=(\partial/\partial t+\gamma+v\mathbf{e}
_d\cdot\nabla)$. As $\det(\mathrm{A}-\gamma\Rm^{\pm1})=\prod_d A_d-\gamma^z$ for $\mathrm{A}=
\mathrm{diag}(A_d)$, the marginal density $\rho=\sum_dP_d$ satisfies
\begin{equation}
\left[\prod_{d\in\mathcal{D}}\left(\frac{\partial}{\partial t}+\gamma+v\,\mathbf{e}_d\cdot\nabla
\right)-\gamma^z\right]\rho(\rb,t)=0,
\label{eq:cyclicPDE}
\end{equation}
the \textit{telegrapher's equation\/} for $z=2$ in one dimension. Three planar directions, $z=3$,
lead to Bessel functions of third order \cite{di2002exact,orsingher2002bessel,leorato2004bose},
and minimal motions in $\R^d$ ($d$ here exceptionally denotes spatial dimension) to hyper-Bessel
functions \cite{lachal2006minimal}, with extensions to orthogonal \cite{leorato2003alternating, 
cinque2023stochastic}, three-dimensional \cite{orsingher2004cyclic}, correlated
\cite{marchione2024planar} and general \cite{lachal2006cyclic} settings. The planar motion in a
vortex \cite{orsingher2025planar} is a continuous-time version of the coin \eqref{eq:geneal_coin}.

\section{Memory in the environment and in the path}
\label{sec:environment}

In all walks considered so far, the coin is identical in every site, so that in Fourier space
equation \eqref{eq:CRW_time_evolution} reduces to the $z\times z$ matrix $\hat{\mathrm{S}}(
\mathbf{q})\C$, and the transport is encoded in the parameter $\lambda$. Once the coin depends
on position or history, this reduction is lost, and exact results of this section instead rest
on invertibility, mappings onto percolation hulls, the combinatorics of Euler circuits and
limit theorems for random environments.

\subsection{Lorentz lattice gases}

In \textit{Lorentz lattice gases\/} \cite{ruijgrok1988deterministic}, a particle moves
ballistically along lattice bonds and typically every site carries a deterministic coin
from the $\C_{\rb}$ symmetry group of the square $D_4$, which acts on the direction of
the walker as an orthogonal matrix $\E\C_{\rb}=\cm_{\rb} \E$, see figure \ref{fig:environment}(a).
Rotating sites ("rotators")
act as $\Rm$ or $\Lm$, with $\cm=\pm\eps$, and thus are chiral. Diagonally reflecting sites
("diagonal mirrors") reflect the internal degree about their own axis, i.e., they turn $d={
\rightarrow}$ into $d={\uparrow}$ and $d={\uparrow}$ into $d={\rightarrow}$. Yet diagonal
sites are achiral, as each is invariant under a reflection of the lattice. Note that this
diagonal mirroring operation is \textit{not\/} the same as the backscattering coin $\Bm$,
which corresponds to a rotation by $\pi$ with $\cm=-\one$. As all these operations are
permutations, the time evolution operation $\mathrm{S}[\sum_{\rb}|\rb)(\rb|\otimes\C_{\rb}]$
is a permutation of the states $|\rb,d)$: trajectories never merge, and each trajectory is
either closed or unbounded. A square lattice equally covered by both diagonal mirrors closes
all trajectories on critical percolation hulls \cite{ziff1991lorentz,cohen1995new}, whereas
partially covered mirror lattices show anomalous diffusion \cite{kong1989anomalous,cohen1995new}.
Rotators instead trap the particle at all concentrations except on critical lines and show the
same dynamical exponent \cite{cohen1995new,wang1995diffusion,cao1997scaling}. Averaging over
coins at every step, which is the Boltzmann approximation of Lorentz-gas kinetic theory
\cite{ernst1988lorentz,vanvelzen1991deterministic}, yields $\cm=\sum_s\phi_s\cm_s$, where
$\phi_s$ denotes the concentration of species $s$. As mirrors and backscatterers only contribute
symmetric matrices, $D_\perp$ here stems from rotators alone. Thus specified to rotators,
$\lambda=1-\phi_R-\phi_L+\mathrm{i}(\phi_R-\phi_L)$\footnote{At full rotator coverage this is
exactly Gillis's coin \cite{gillis1960random} with $\alpha=\phi_R-\phi_L$ or similarly the CRW
of Ref.~\cite{wojcik2026chiral} with $\chi=(\phi_R-\phi_L)$.}, and from equation
\eqref{eq:odd_diff_general_coin} it follows that $D_\perp\propto(\phi_R-\phi_L)$. Mixtures
with backscatterers, with rotators \cite{gunn1985percolation} or with stochastic scatterers 
\cite{acedo1999diffusion}, connect transport to percolation; the latter show a diffusive
percolation threshold at backscatterer fraction $2/3$ on the square lattice in the
low-density limit \cite{acedo1999diffusion}. Rotator gases; an achiral walker among chiral
scatterers, and the disordered CRW of Ref.~\cite{uhlig2026chirality}; a chiral walker among
achiral obstacles, are thus two implementations of the same concept.

\subsection{Kinetic walks and self-modifying environments}

We now consider coins $\C_{\rb}$ that change when the walker visits the specific site and thus
create an environment, which stores the history of the walk. In the \textit{hull-generating
walk\/} of Refs.~\cite{ziff1984generation,ziff1989hull}, each coin is drawn at the first visit
and then frozen: the walker on the honeycomb lattice turns left or right according to whether
the hexagon ahead is occupied, with probability $p$, or empty. This constitutes a "quenched
disorder" sampled on the fly; historically this approach was used to create percolating clusters
numerically. The walk thus produces a Lorentz gas \cite{ziff1991lorentz} of rotators that turn
by $\pm\pi/3$, with an excess of left over right turns $q_L-q_R=2p-1=2(p-p_c)$, where $p_c=1/2$
is the threshold of site percolation on the triangular lattice. At $p_c$, the walk percolates
with a scaling limit corresponding to the Schramm-Loewner evolution SLE$_6$
\cite{schramm2000scaling,camia2007critical}, corresponding to the fractal dimension $7/4$
\cite{beffara2008dimension}. This walk dimension is obtained exactly for the CRW on critical
percolation clusters in Ref.~\cite{uhlig2026exact} in the fully chiral limit, where
the walker traces the hull. Chirality thus turns the normally subdiffusive random walk on
fractal structures superdiffusive, with a different, yet exact MSD exponent
\cite{uhlig2026exact}. For $p\neq p_c$, the walk traces the perimeters of off-critical
clusters \cite{weinrib1985new}, and thus plateaus eventually. If each visit instead inverts the
coin, $\Rm\leftrightarrow\Lm$, the walk is a \textit{flipping-rotator gas}
\cite{ruijgrok1988deterministic}, studied also on the triangular \cite{kong1991diffusion} and
honeycomb \cite{wang1995diffusion} lattices, which includes Langton's ant
\cite{langton1986studying}, a famous cellular automaton that turns right on white and left on
black squares and flips their colour, and whose trajectories are unbounded
\cite{bunimovich1992recurrence}, see also figure \ref{fig:environment}(b).

If each visit advances the coin, the walk is a \textit{rotor walk}, or \textit{Eulerian walker\/}
\cite{priezzhev1996eulerian}: every site stores an orientation $d_{\rb}$, which rotates the
arriving walker by $\Rm$ before leaving along this new direction. The next step thus depends
not on the walker's previous step but on the previous departure from the same site: successive
departures from a site are rotated by $\Rm$, similar to successive steps in the CRW at $\chi=1$.
The same rotation that closes the CRW on a single elementary square makes the rotor walk cover 
its surroundings: on a finite lattice, it locks into an Euler circuit that traverses every
bond once in each direction \cite{priezzhev1996eulerian}. On $\Z^2$, from independent uniform
initial rotors, the visited region grows with radius $\propto t^{1/3}$ and appears circular
\cite{priezzhev1996eulerian,kapri2009asymptotic}; the range exponent $2/3$ and the convergence
to a convex limit shape are now proven \cite{bou2026eulerian}. Specifically, stochastic rotation
rules restore random-walk behaviour \cite{kapri2009asymptotic}.

\subsection{Random odd environments}

A walker without internal states can also break detailed balance if its jump rates $k_{\rb\rb'}$
from $\rb$ to a neighbour site $\rb'$ vary in space. For doubly-stochastic rates, $\sum_{\rb'}k_
{\rb\rb'}=\sum_{\rb'}k_{\rb'\rb}$, as in the environments considered by Kozma and T\'oth
\cite{kozma2017central}, the uniform density is stationary, and the master equation reads
$\partial\rho(\rb,t)/\partial t=\sum_{\rb'}k_{\rb'\rb}[\rho(\rb',t)-\rho(\rb,t)]$. The rates
split up into a symmetric part $s_{\rb\rb'}= (k_{\rb\rb'}+k_{\rb'\rb})/2$ and an antisymmetric 
part $a_{\rb\rb'}=(k_{\rb\rb'}-k_{\rb'\rb})/2$. Having no net outflow on the square lattice is
equivalent to
\begin{equation}
a_{\rb,\rb+\ell\mathbf{e}}=\psi\left(\bar{\rb}-\ell\eps\mathbf{e}/2\right)-\psi\left(\bar{\rb}
+\ell\eps\mathbf{e}/2\right),
\label{eq:curl}
\end{equation}
where $\psi$ is the stream function that is defined on the centres of the elementary squares of
the lattice, $\bar{\rb}=\rb+\ell\mathbf{e}/2$, and $\bar{\rb}\pm\ell\eps\mathbf{e}/2$ are the
centres of the two neighbouring cells. Detailed balance, as in observation (ii) above, holds if
and only if $a=0$, i.e., if $\psi$ is constant; otherwise, the increments are correlated through
the quenched environment, which provides the memory required by the hypothesis.

For a constant $s_{\rb\rb'}=s$ and a slowly varying $\psi$, $a_{\rb,\rb+\ell\mathbf{e}}\simeq-
\ell\,(\eps\mathbf{e})\cdot\nabla\psi$, and the master equation becomes $\partial\rho/\partial
t=s\ell^2\nabla^2\rho-\mathbf{u}\cdot\nabla \rho$ with $\mathbf{u}=2\ell^2\eps\nabla\psi$.
Because of
\begin{equation}
\nabla\cdot[(D_\parallel\one+D_\perp(\rb)\eps)\nabla\rho] = D_\parallel\nabla^2 
\rho-\mathbf{u}\cdot\nabla\rho,
\label{eq:stream}
\end{equation}
where $\mathbf{u}=\eps\,\nabla D_\perp(\rb)$, this corresponds to space-dependent odd diffusion
$D_\perp=2\ell^2\psi$, generated by the stream function of the rates, also $D_\parallel=s\ell^2$
in equation (\ref{eq:stream}). The mean of $D_\perp$ thus drops out of the density dynamics and
enters only through flux boundary conditions \cite{kalz2022collisions}, whereas its spatial
variations affect the density \cite{abdoli2020nondiffusive}, and act, for a random $D_\perp$,
as a random incompressible flow. A central limit theorem holds if the stream function is
stationary and square integrable~\cite{kozma2017central,toth2026central}.

\subsection{Memory in the path: spiral self-avoiding walks}

Finally, \textit{spiral self-avoiding walks\/} \cite{privman1983spiral} store the memory
in the path. They step straight ahead, $\I$, or turn in one sense, e.g., $\Rm$, only, but
self-avoidance replaces the coin by a constraint on the entire past, see also figure
\ref{fig:environment}(d). The exact solution
\cite{blote1984spiralling}, refined in Refs.~\cite{guttmann1984number,brak1998anisotropic},
shows that the number of distinct $N$-step spiral self-avoiding walks grows as $\propto N^{-7/4}
\exp(2\pi\sqrt{N/3})$, a growth slower than exponential. Moreover, the exact solution shows
that $N$-step walks have an end-to-end distance that scales as $\propto N^{1/2}\ln N$. Full
chirality combined with self-avoidance thus yields neither diffusive nor localised behaviour. 

\begin{figure*}
\includegraphics[width=\textwidth]{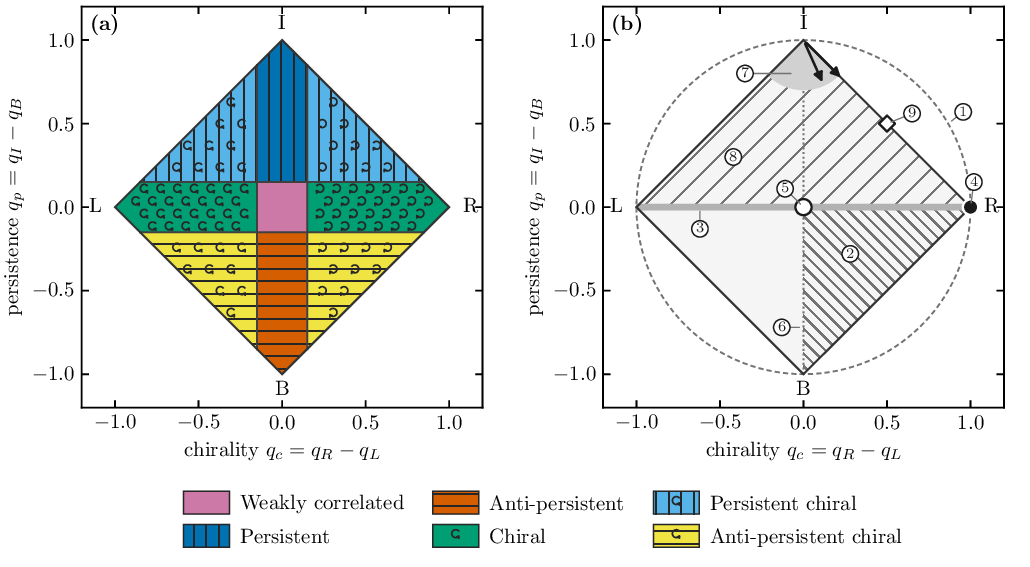}
\caption{Persistence-chirality diamond of the general coin matrix \eqref{eq:geneal_coin}. The
diamond shape represents the normalisation constraint. The vertices I, L, B and R are the pure
persistent, left-turning, reversing and right-turning coins. (a) Sketch of the dynamic
behaviours that the active CRW can exhibit. (b) Literature walks placed by their polar
eigenvalue $\lambda=q_p+\mathrm{i}q_c$. Generic four-turn coins fill the whole diamond
\cite{henderson1984correlated,chen1994general,gilbert2010diffusion,sherf2026active}, as do
Lorentz gases that mix rotators and backscatterers in the Boltzmann approximation
\cite{gunn1985percolation,acedo1999diffusion}. (1) Walks with continuous turning angles fill
the unit disk (dashed), of which only the diamond is reachable on the square lattice
\cite{kareiva1983analyzing,bracher2004eigenfunction,sadjadi2015persistent,larralde1997transport}.
(2) CRW in annealed disordered environment \cite{uhlig2026chirality}. (3) $q_p=0$, chiral walks:
Gillis' walk \cite{gillis1960random}, the rotator Lorentz gas \cite{ruijgrok1988deterministic,
cohen1995new}, flipping rotators such as Langton's ant \cite{langton1986studying,
bunimovich1992recurrence}, as well as the CRW \cite{wojcik2026chiral}. (4) Fully chiral vertex R:
CRW at $\chi=1$ \cite{wojcik2026chiral}, Gillis' walk at $\alpha=1$ \cite{gillis1960random}, as
well as the topological chiral random walker at $D_r=0$, $\omega=1$ \cite{osat2026topological}.
(5) $\lambda=0$: the normal random walk \cite{henderson1984correlated}. (6)~$q_c=0$: achiral
persistent walks \cite{patlak1953random}. (7) Continuous-time walks, $\lambda=1-(\Gamma-\mathrm{i}
\Omega)\Delta t$ with $\Delta t\to0$, sit at I and leave it within the wedge $|\kappa|\leq1$
(arrows along $\kappa=\Omega/\Gamma$): chiral random motion \cite{hargus2021odd}, chiral and
orthogonal run-and-tumble walks \cite{mallikarjun2023chiral,angelani2022orthogonal} and the
weak-turning limit of the active CRW \cite{sherf2026active}; cyclic random motions with four 
directions \cite{leorato2003alternating} run along the edges I-R and I-L. (8) In the Boltzmann
approximation, rotator Lorentz gases with empty sites fill the upper triangle I-L-R
\cite{ruijgrok1988deterministic,cohen1995new,vanvelzen1991deterministic}; meeting the chiral
walker among achiral obstacles \cite{uhlig2026chirality} of (2) on the line $q_p=0$. (9) Spiral
self-avoiding walks: the coin $(\I+\Rm)/2$, at $\lambda=(1+\mathrm{i})/2$, conditioned on
self-avoidance \cite{privman1983spiral,blote1984spiralling}.}
\label{fig:literature}
\end{figure*}

\section{Conclusion and outlook}
\label{sec:outlook}

We organised chiral lattice walks around the one-step operator $\Sm(\one\otimes\C)$ following
the hypothesis that chirality requires at least a two-step memory, stored in the walker, in
the environment, or in the path. For isotropic coins, the diffusion tensor depends only on
the polar eigenvalue $\lambda$, which places our coin models \cite{wojcik2026chiral,
uhlig2026chirality,sherf2026active} and the historical walks in a common, umbrella picture.
A graphical overview is provided in figure \ref{fig:literature}. Across all families, full
chirality in discrete time closes orbits, partial chirality yields optima, and edges carry
currents.

Current work extends the content of our review in multiple directions. The first is the case
of many walkers. In mass-conserving jump processes with a preferred rotational sense, 
independent single-bond transfers yield a diagonal diffusion tensor, and the minimal chiral
move is an L-shaped transfer across two orthogonal bonds \cite{das2026odd}, the many-body
counterpart of the two-step memory of section \ref{sec:crw}. Odd diffusion then generates
density correlations decaying as $r^{-4}$, the mobility remains related to the steady-state 
fluctuations through the symmetric part of $\D$ only, and reflecting walls carry a steady
edge current \cite{das2026odd}. The second direction concerns interfaces. In a chiral active
lattice gas with biased local rotations, the system phase-separates and condensate interfaces
carry persistent, unidirectional edge currents \cite{wang2026edge}. The mean-field theory of
a chiral Ising model, the chiral active Model B, shows arrested coarsening and details how
the microscopic rotational bias generates tangential currents localised at interfaces
\cite{blom2026mean}. Both works thus investigate a many-body analogue of the edge transport
of section~\ref{sec:crw}. The third direction relates to active lattice walkers under
confinement and in a flow. Aligning particles form clogs and bands
\cite{calascibetta2024effects} and invite an interpretation of the coin language of the
active CRW \cite{sherf2026active}. On the mathematical side, the recent proofs for rotor
walks and divergence-free environments \cite{bou2026eulerian,toth2026central} discussed in
section \ref{sec:environment} show that memory stored in the environment is becoming tractable.

The lattice perspective developed here suggests a direct experimental route using programmable
active particles. If continuing, reversing, and turning left or right can be prescribed
independently, the coin of equation \eqref{eq:geneal_coin} could be implemented directly and the
persistence-chirality diamond of figure \ref{fig:literature} explored experimentally, including
the strongly chiral and anti-persistent regimes that have no simple continuum counterpart.

From a random-search perspective \cite{targetbook}, it will be of interest to combine the CRW
and other lattice random walk models with the mechanism of stochastic resetting \cite{evansprl,
evansreview}. In stochastic resetting, a walker is "reset" to either its starting position
("complete resetting") \cite{chechkin,yael,shlomi,yael1,ciliberto,ewa,ewa1,trifce,trifce1}
or to a fraction of its value ("partial resetting") \cite{marcus,costantino,costantino1}.
As resetting is known to optimise mean first-passage times under certain conditions, it should
be analogously possible to improve the spatial exploration of chiral lattice random walks by
stochastic resetting.

We finish with some open problems, which consecutively become more speculative, but which
appear natural from the lattice perspective taken in this review. First, it is unclear whether
the hypothesis that chirality requires memory can be proven, or refuted by a counterexample,
when the memory resides in a heterogeneous environment or in the path rather than in the
walker. Second, the CRW has a renewal structure, as every mixing event erases its orientation
and the walk is deterministic in between; this suggests closed-form propagators, also in
bounded domains, which would underpin a full theory of chiral first passage beyond the optima
found so far \cite{mallikarjun2023chiral,uhlig2026chirality}. Third, chiral walkers among 
chiral scatterers would combine internal and environmental memory, and it is open whether
the two kinds of chirality add up or compete. Fourth, an isotropic coin violates detailed
balance exactly when $q_L\neq q_R$, i.e., under the same condition that renders it odd,
which raises the question whether dissipation bounds the oddness, and how it distinguishes
active from Onsager-Casimir chirality. Fifth, it is not known whether the chirality of rotor
and flipping-rotator rules, which store their memory in the sites, leaves an odd signature
in transport. Finally, beyond the square lattice, an isotropic diffusion tensor cannot be
odd in three dimensions, so that chirality there requires a preferred axis, such as a field
or the helical axis of the walker, while on networks and on curved or defected lattices the
number of directions changes from site to site and the coin is no longer a polynomial in a
single rotation. Both situations have not been investigated to our knowledge.

\begin{acknowledgments}
E. K. and R. M. acknowledge fruitful discussion at the "39th Marian von Smoluchowski Symposium
on Statistical Physics" in Rome 2026, which stimulated this review. R.M. acknowledges
financial support by the German Research Foundation (DFG, grants no. ME 1535/22-1 and no.
318763901, within the CRC 1294 "Data Assimilation", project B10). Y.R. and N.S. acknowledge 
support from the Israel Science Foundation (grants no. 385/21 and no. 2013/25).
\end{acknowledgments}

\appendix

\section{Matrix representation of the coin operators}
\label{sec:matrices}

The elementary operators of the CRW are the identity $\mathrm I$, the right-chiral $\mathrm R$,
left-chiral $\mathrm L$, reversal $\mathrm B$, and mixing (randomisation) $\mathrm M$ operators.
Representing the basis of the internal orientation by the vectors $\rket{\rightarrow}=(1,0,0,0)
^\mathrm{T}$, $\rket{\downarrow}=(0,1,0,0)^\mathrm{T}$, $\rket{\leftarrow}=(0,0,1,0)^\mathrm{T}$,
and $\rket{\uparrow}=(0,0,0,1)^\mathrm{T}$, allows us to write the operators as $4\times4$
matrices,
\begin{equation}
\begin{split} &  \mathrm I=\begin{pmatrix}1&0&0&0\\
0&1&0&0\\
0&0&1&0\\
0&0&0&1\\\end{pmatrix},
~
\mathrm L=\begin{pmatrix}
0&1&0&0\\
0&0&1&0\\
0&0&0&1\\
1&0&0&0\\
\end{pmatrix},
~
\ \mathrm R=\begin{pmatrix}
0&0&0&1\\
1&0&0&0\\
0&1&0&0\\
0&0&1&0\\
\end{pmatrix},\\
& \mathrm B=\begin{pmatrix}
0&0&1&0\\
0&0&0&1\\
1&0&0&0\\
0&1&0&0\\
\end{pmatrix},
~
\mathrm M=\frac{1}{4}\begin{pmatrix}
1&1&1&1\\
1&1&1&1\\
1&1&1&1\\
1&1&1&1\\
\end{pmatrix}.
\end{split}
\end{equation}
The general coin of equation \eqref{eq:geneal_coin} is a combination of these matrices given by
\cite{sherf2026active} 
\begin{equation}
\mathrm C=q_I\mathrm I+q_R\mathrm R+q_L\mathrm L+q_B\mathrm B+q_M\mathrm M,
\end{equation}
where $q_I$, $q_R$, $q_L$, $q_B$ and $q_M$ are the probabilities of choosing the respective
coins. These probabilities are bounded in $[0,1]$ and obey the normalisation $q_I+q_R+q_L+q_B
+q_M=1$. The eigenvalues of $\mathrm C$ are given by
\begin{subequations}
\begin{align}
\lambda_0&=1,
\label{eq:lambda0}\\
\lambda_1&=\left(q_I-q_B\right)+\mathrm{i}\left(q_R- q_L\right)\equiv q_p+\mathrm{i}\,q_c,
\label{eq:lambda1}\\
\lambda_2&=\left(q_I+ q_B\right)-\left(q_R+ q_L\right)\equiv q_n,
\label{eq:lambda2}\\
\lambda_3&=\left(q_I-q_B\right)-\mathrm{i}\left(q_R-q_L\right)=\overline{\lambda_1}.
\label{eq:lambda3}
\end{align}
\end{subequations}
For a detailed discussion of these eigenvalues, see Ref.~\cite{sherf2026active}.

\end{document}